\documentclass[aps,showpacs,groupedaddress,floatfix,twocolumn,nofootinbib,reprint,nopacs
]{revtex4-1}

\usepackage{graphicx}
\usepackage{bbm}
\usepackage{amsmath}
\usepackage{amssymb}
\usepackage{braket}

\DeclareGraphicsExtensions{.eps, .eps, .jpg, .png}

\def\bk{{\bf k}}
\def\bq{{\bf q}}
\def\bx{{\bf x}}

\begin{document}

\begin{flushright}  ACFI-T15-17  \end{flushright}

\title{On the electric charge of the observable Universe}

\author{Cody Goolsby-Cole} 

\email[]{cgoolsby@physics.umass.edu}

\author{Lorenzo Sorbo} 

\email[]{sorbo@physics.umass.edu}

\affiliation{Amherst Center for Fundamental Interactions, Department of Physics, University of Massachusetts, Amherst, MA 01003}


\begin{abstract}
Light fields get large scale fluctuations during inflation. If some of them are electrically charged, then large scale fluctuations of the electric charge will be generated. As a consequence, any finite portion of the Universe, including our observable one, will carry a net electric charge. This fact does not require any form of breaking of the gauge symmetry at any time.  We discuss under which conditions such a charge is maintained until the end of inflation, and we estimate its expected magnitude  both in the case of charged fermions and of charged scalars.  While one charged fermion species yields a charge density that is several orders of magnitude below the observational constraints, extremely light charged scalars can  exceed those constraints.

\end{abstract}

\maketitle

\section{Introduction}%

The conservation of electric charge is one of the best established, least questioned laws of physics~\cite{pdg}. While scenarios  where charge is not conserved have been proposed (see below for an incomplete list), these correspond to exotic situations, and  the charge of the Universe is usually assumed to vanish. 

The present work focuses on the fact that, even if electric charge is exactly conserved as a global quantity, during inflation with Hubble parameter $H$ large scale charge fluctuations are generated if there exist charged particles with mass $m\lesssim H$. As a consequence, even if the entire Universe is  electrically neutral, any finite portion (including our observable one) of it can have a net charge. We will estimate the typical magnitude of the average charge density ${\mathcal Q}_R$ in a volume of radius $R$ by computing its variance right after inflation.

The constraints on the electric charge density ${\mathcal Q}_0$ of the Universe are tight, ${\mathcal Q}_0\lesssim 10^{-26} n_B$, where the $n_B$ is the number density of baryons~\cite{Caprini:2003gz} (see also~\cite{Orito:1985cf,Masso:2002vh} for previous analyses that did not account for the large conductivity of the primordial plasma). For charged massive {\em fermions} of mass $m={\cal O}(H)$ we will find a charge density that is orders of magnitude smaller, ${\mathcal Q}_0\lesssim 10^{-33}n_B$. The charge density in {\em scalars with $m={\cal O}(H)$} will be only two or three orders of magnitude larger. However, the  charge density accumulated in {\em very light charged scalar} particles can be much larger, and can exceed by several orders of magnitude the bounds of~\cite{Caprini:2003gz} in the limit of a massless scalar species. 

One might worry that the electric field produced by the charge inhomogeneities during inflation can oppose charge separation or annihilate the charges via Schwinger effect. As we will see, this is typically not the case, even if in some instances the Schwinger effect can be relevant.

The idea that the Universe might carry a net electric charge dates back to the work of Lyttleton and Bondi~\cite{Lyttleton:1959zz}, who assumed ${\mathcal Q}_0/n_B\simeq 10^{-18}$ to explain the recession of distant galaxies, while in the '60s Alfv\'en and Klein~\cite{alfven} considered a cosmology where charge separation would play a central role. The possibility of a charge imbalance, analogous to that of~\cite{Lyttleton:1959zz}, but confined to dark matter was discussed more recently in~\cite{Kaloper:2009nc}.  References~\cite{Ignatiev:1978xj,Langacker:1980kd,Enqvist:1988br,Dolgov:1990sg} considered the generation of a net charge caused by the spontaneous breaking of the electromagnetic gauge symmetry  (used in~\cite{Dolgov:1993mu} to generate cosmological magnetic fields), and~\cite{Dolgov:2006bc} has shown that the same effect is produced by a photon mass. The authors of~\cite{Dubovsky:2000av} discussed the possibility that electric charge is not conserved in brane world models.  Closer to our work, a massive charged scalar during inflation was discussed in~\cite{Calzetta:1997ku}, whose focus, however, was on the generation of magnetic fields. The system of~\cite{Calzetta:1997ku} was reanalyzed in~\cite{Giovannini:2000dj}, where the current charge density of the Universe was also estimated in the case of a massless charged scalar. A charged curvaton was considered in~\cite{D'Onofrio:2012qy}, where it was argued that the charge density should not survive until the end of inflation because of Schwinger pair production. In~\cite{Kobakhidze:2015zka}, a mechanism analogous to ours, with the electric charge replaced by the baryon number, was proposed to produce the observed baryon asymmetry of our Universe. As we discuss below, our results differ significantly from those of~\cite{Giovannini:2000dj,D'Onofrio:2012qy,Kobakhidze:2015zka}.

\section{Charge density during inflation}%

We define the average charge density ${\mathcal Q}_R$ in a volume of radius $R$ as
\begin{equation}
{\mathcal Q}_R\equiv\int \frac{d^3\bx}{(\sqrt{\pi}\,R)^3}\,e^{-\bx^2/R^2}\,{\mathcal Q}(\bx)\,,
\end{equation}
where ${\mathcal Q}(\bx)$ is the charge density operator for the form of matter under consideration. Since electric charge is conserved, and any initial charge density will be rapidly driven to zero by inflation, $\langle {\mathcal Q}_R\rangle=0$. However the {\em variance} of ${\mathcal Q}_R$ will not vanish and its square root will give the typical size of the charge density in a sphere of radius $R$. If we define the charge power spectrum $P_{\mathcal Q}(k)$
\begin{equation}
\langle {\mathcal Q}(\bk)\,{\mathcal Q}(\bk')\rangle\equiv \frac{2\pi^2}{k^3}\,\delta(\bk+\bk')\,P_{\mathcal Q}(k)\,,
\end{equation}
with ${\mathcal Q}(\bk)\equiv \int\frac{d^3\bx}{(2\pi)^{3/2}}e^{-i\bk\bx}{\mathcal Q}(\bx)$, it is then straightforward to prove that
\begin{equation}\label{qrqk}
\langle {\mathcal Q}_R^2\rangle=\int\frac{dk}{k}\,P_{\mathcal Q}(k)\,e^{-k^2R^2/2}\,.
\end{equation}
We will be caring about the limit of large $R$, so that we will need to compute only $P_{\mathcal Q}(k)$ for $k\to 0$.

Since we are ultimately interested in the number density of (anti)particles, it is crucial to work with normal-ordered operators to avoid divergences. The rigorous way of normal-ordering operators in a time-dependent background is to compute the Bogolyubov coefficients, as we will discuss below and in Appendix~\ref{app:bogos}. To our knowledge, this is the first work where the charge variance in Dirac fermions is computed at the end of inflation. On the other hand, the charge variance of complex scalar fields during inflation was considered in~\cite{Giovannini:2000dj,D'Onofrio:2012qy,Kobakhidze:2015zka}, where however normal ordering was not taken into account. Because of this, our results differ significantly from those of~\cite{Giovannini:2000dj,D'Onofrio:2012qy,Kobakhidze:2015zka}. We believe that our analysis, which gives unambiguously finite results, is the appropriate one for this problem.

\subsection{Fermions}

Let $\psi$ be a fermion with mass $m$ in a FRW Universe with scale factor $a(\tau)$, where $\tau$ is conformal time. The canonically normalized field $\Psi\equiv a^{3/2}\,\psi$ obeys 
\begin{equation}\label{dirac_eq}
\left(i\,\gamma^\mu\,\partial_\mu-m\,a\right)\,\Psi=0
\end{equation}
that we solve by decomposing
\begin{align}
\Psi(\bk,\,\tau)=\sum_{r=\pm 1}\left[u_r(\bk,\,\tau)\,a_r(\bk)+v_r(\bk,\,\tau)\,b^\dagger_r(-\bk)\right]\,,
\end{align}
with (using the conventions of~\cite{Nilles:2001fg})
\begin{align}
&u_r(\bk,\,\tau)=\frac{1}{\sqrt{2}}\left(
\begin{array}{c}
U_+(k,\,\tau)\,\psi_r(\hat\bk) \\
r\,U_-(k,\,\tau)\,\psi_r(\hat\bk)
\end{array}
\right)\,,\nonumber\\
&v_r(\bk,\,\tau)=\frac{1}{\sqrt{2}}\left(
\begin{array}{c}
V_+(k,\,\tau)\,\psi_r(\hat\bk) \\
r\,V_-(k,\,\tau)\,\psi_r(\hat\bk)
\end{array}
\right)\,,
\end{align}
where $\psi_r$ is an eigenfunction of the helicity operator with eigenvalue $r/2$. The equations of motion read
\begin{equation}\label{dirac_eq_mom}
U_\pm'=-i\,k\,U_\mp\mp i\,m\,a\,U_\pm\,.
\end{equation}
Given that the system is invariant under charge conjugation, we have $V_+=-U_-^*$, $V_-=U_+^*$. Moreover, the normalization  $|U_+|^2+|U_-|^2=2$ is preserved by the equations of motion.

In a de Sitter geometry $a(\tau)=-(H\,\tau)^{-1}$, eqs.~(\ref{dirac_eq_mom}) are solved by
\begin{equation}
U_\pm=\sqrt{\frac{-\pi k\tau}{2}}\,e^{\pm\frac{\pi\,m}{2\,H}}\,{\mathrm H}^{(1)}_{\frac{1}{2}\mp i\frac{m}{H}}(-k\tau)\,,
\end{equation}
where ${\mathrm H}^{(1)}_\nu(x)$ denotes the Hankel function of the first kind.

In order to compute the renormalized two point function of the charge operator we compute the Bogolyubov coefficients for this system. To do so we decompose $\Psi(\bk,\,\tau)$ on a different set of creation/annihilation operators $\tilde{a}^{(\dagger)}_r(\bk,\,\tau)$,  $\tilde{b}^{(\dagger)}_r(-\bk,\,\tau)$ and mode functions $\tilde{U}_\pm(k,\,\tau)$ that are the adiabatic solutions of eqs.~(\ref{dirac_eq_mom}) 
\begin{align}
\tilde{U}_\pm=\left(1\pm\frac{m\,a}{\sqrt{k^2+m^2\,a^2}}\right)^{1/2}\,e^{-i\int {\sqrt{k^2+m^2\,a^2}}\, d\tau}
\end{align}
and are linearly related to the functions $U_\pm$ by
\begin{align}
&U_+(\bk,\,\tau)=\alpha(\bk,\,\tau)\,\tilde{U}_+(\bk,\,\tau)-\beta(\bk,\,\tau)\,\tilde{U}_-^*(\bk,\,\tau)\nonumber\\
&U_-(\bk,\,\tau)=\alpha(\bk,\,\tau)\,\tilde{U}_-(\bk,\,\tau)+\beta(\bk,\,\tau)\,\tilde{U}_+^*(\bk,\,\tau)\,.
\end{align}

By definition, during adiabatic evolution, $\omega'\ll \omega^2$, the Bogolyubov coefficients $\alpha(\bk,\,\tau)$ and $\beta(\bk,\,\tau)$ are constant, and  the occupation number for modes with momentum $\bk$ is given by $\langle 0|\tilde{a}(\bk)^\dagger\tilde{a}(\bk)|0\rangle=|\beta(\bk)|^2$, where the vacuum $|0\rangle$ is annihilated by the $a_r(\bk)$, $b_r(\bk)$ operators.

For modes with $k\ll m\,a$ the adiabaticity condition reads $a'/a^2\ll m$. During inflation this condition is not satisfied for the fermions with $m\lesssim H$ we are considering, but it is after inflation ends, when the Hubble parameter $a'/a^2$ decreases. Therefore to compute the Bogolyubov coefficients we join the inflationary period to a radiation dominated one with $a(\tau)=H\,\tau+2$ for $\tau>-1/H$. The equations of motion for $U_\pm$ during radiation domination  can be solved in terms of parabolic cylinder functions and yield the final value of the Bogolyubov coefficients,  whose main feature is that $k^3\,|\beta(k)|^2$ is peaked at $k\simeq m$. Their explicit expression, which is long and not very illuminating, will not be presented here. 

\begin{figure}
\centering
    \includegraphics[width=.4\textwidth]{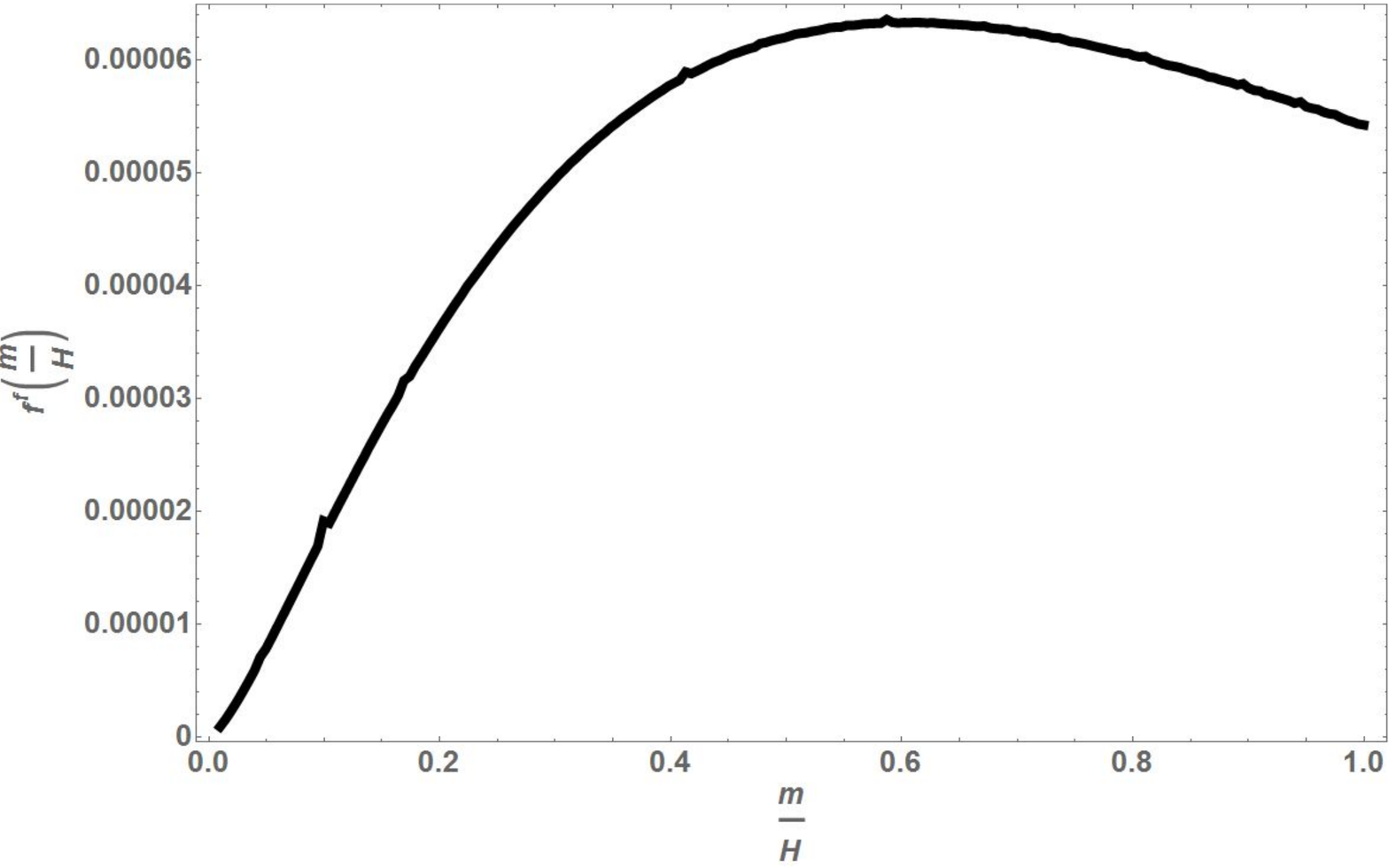}
    \caption{Numerical evaluation of the function $f^f(m/H)$ appearing in eq.~(\ref{fermion_final}) and giving the overall amplitude of the charge power spectrum for fermions of mass $m$ during inflation with Hubble parameter $H$.}\label{fmh}
\end{figure}

The normal ordered (in terms of the tilded operators) two point function of the charge is
\begin{align}\label{two_q_zero}
\langle :{\mathcal Q}(\bk)&{\mathcal Q}(\bk'):\rangle=e^2\,\int \frac{d^3\bx\,d^3{\bf y}}{(2\pi)^3}\,e^{-i\bk\bx-i\bk'{\bf y}}\nonumber\\
&\times \langle : \Psi^\dagger(\bx,\,\tau)\Psi(\bx,\,\tau)\,\Psi^\dagger({\bf y},\,\tau)\Psi({\bf y},\,\tau):\rangle\,,
\end{align}
that, in the limit $\bk,\,\bk'\to 0$, gives
\begin{align}\label{fermion_final}
P^f_{\mathcal Q}(k\to 0)&=-\frac{e^2\,k^3}{2^2\,\pi^5}\int{d^3\bq}\left|\beta\right|^2\equiv -e^2\,k^3\,H^3\,f^f\left(\frac{m}{H}\right)\,,
\end{align}
where the function $f^f(m/H)$, plotted in figure~\ref{fmh}, shows that, for $m\sim H$,  $P_{\mathcal Q}^f(k)\sim 10^{-5}\,e^2\,k^3\,H^3$.

\subsection{Scalars}\label{sub:scalars}

The case of a complex scalar is treated similarly, but, due to the absence of Pauli blocking, will lead to a richer set of possibilities.  The canonically normalized field $\varphi$ satisfies
\begin{equation} \label{scalar}
\varphi'' + \left( k^2+ m^2\,a^2 - \frac{a''}{a} \right)\,\varphi = 0
\end{equation}
that is solved by decomposing
\begin{align}\label{scalardeco}
 \varphi(\bk,\, \tau) \equiv  \phi(\bk,\, \tau)\, a(\bk) + \phi^*(- \bk, \tau)\, b^\dagger(- \bk)\,,
\end{align}
where the mode functions read
\begin{equation} \label{mode}
\phi(\bk,\, \tau)= \sqrt{\frac{-\pi \tau }{4}} \mathrm{H}^{(1)}_\nu(-k\,\tau),\quad \nu \equiv \sqrt{ \frac{9}{4} - \frac{m^2}{H^2}  }\,,
\end{equation}
where we assume $m<\frac{3}{2}\,H$.  As we did for fermions, we then decompose $\varphi(\bk,\,\tau)$ using a different set of operators $\tilde a(\bk,\,\tau)$ and $\tilde b(\bk,\,\tau)$ and the adiabatic mode functions
\begin{align} \label{admode}
\tilde \phi(k,\,\tau) = \frac{e^{-i \int \omega_k d\tau}}{\sqrt{2\omega_k}} \,,\quad
\omega^2_k\equiv k^2+m^2\,a^2-\frac{a''}{a}.
\end{align}

We then join the solutions during inflation to those obtained during a radiation dominated phase, so that the adiabatic condition is satisfied at late times (see Appendix~\ref{app:bogocalc} for details).

Using the charge operator for scalars 
\begin{equation}
\mathcal{Q}(\bx) = - i e \left[ \varphi^\dagger(\bx,\tau) {\varphi}'(\bx,\tau) -  {\varphi^\dagger}'(\bx,\tau) \varphi(\bx,\tau) \right]\,,
\end{equation}
we obtain
\begin{widetext}
\begin{align}\label{pqscalar}
P_{\cal Q}(k)=  e^2\frac{k^3}{(2\pi)^5} \int \frac{d^3 \textbf{q}}{\omega_{\textbf{k} + \textbf{q}} \omega_\textbf{q}} &\Bigg\{ 2\,|\beta_\textbf{q}|^2 \,|\beta_{\textbf{k} + \textbf{q}}|^2\, (\omega_\textbf{q}^2 + \omega_{\textbf{k} + \textbf{q}}^2) - (\omega_\textbf{q} + \omega_{\textbf{k} + \textbf{q}})^2 \mathrm{Re} \left[ \beta_\textbf{q} \beta^*_{\textbf{k} + \textbf{q}} \alpha_{\textbf{k} + \textbf{q}} \alpha^*_\textbf{q} e^{2i \int \omega_\textbf{q} d\tau - 2i \int \omega_{\textbf{k} +\textbf{q}} d\tau } \right] \nonumber\\ 
& +2 (\omega_{\textbf{k} + \textbf{q}}^2 - \omega_\textbf{q}^2) \left( |\beta_{\textbf{k} + \textbf{q} }|^2 \mathrm{Re}\left[ \beta^*_\textbf{q} \alpha_\textbf{q} e^{-2i \int \omega_\textbf{q} d\tau}\right] - |\beta_{\textbf{q} }|^2 \mathrm{Re}\left[ \beta^*_{ \textbf{k} + \textbf{q} } \alpha_{ \textbf{k} + \textbf{q} } e^{-2i \int \omega_{ \textbf{k} + \textbf{q} } d\tau}\right]  \right)\nonumber\\
 &- (\omega_\textbf{q} - \omega_{\textbf{k} + \textbf{q}})^2 \mathrm{Re}\left[ \beta^*_\textbf{q} \beta^*_{\textbf{k} + \textbf{q}} \alpha_\textbf{q} \alpha_{\textbf{k} + \textbf{q} } e^{-2i \int \omega_\textbf{q} d\tau - 2i \int \omega_{\textbf{k} +\textbf{q}} d\tau }   \right] \Bigg\}
\end{align}
\end{widetext}

The general expression of the Bogolyubov coefficients is rather cumbersome, but in the regime $q \ll H$, $m \lesssim H$, which is of interest for us, it simplifies to~(see appendix~\ref{app:bogocalc})
\begin{align}\label{bogosc}
|\beta_\bq|^2\simeq \left\{
\begin{array}{cc}
.5\times \sqrt{H/m}\times\left(H/q\right)^{2\,\nu}\,, & q\lesssim \sqrt{m\,H}\,,\\
H^4/(4\,q^4)\,, & \sqrt{m\,H}\lesssim q\lesssim H\,.
\end{array}
\right. 
\end{align}
For nonrelativistic massive scalars the phase $\int \omega\, d\tau\simeq m\,H\,\tau^2/2$ (remember that $a(\tau)\simeq H\,\tau$ well after the end of inflation) oscillates rapidly after inflation, so that the second and third lines of equation~(\ref{pqscalar}) can be neglected. Also, one can take $\bk\to 0$ in the first line of that equation since, as we will see, one obtains a finite result. As a consequence, for  scalars with a mass that is large enough, using the relation $|\alpha_\bq|^2-|\beta_\bq|^2=1$, the charge power spectrum can be written in the simple form
\begin{align}\label{pscalmassive}
&P_{\cal Q}(k)=-\frac{e^2}{2^3\pi^5}\,k^3\,\int d^3 \bq \,|\beta_\bq|^2 \nonumber\\
&\simeq -\frac{3\,e^2}{8\,\pi^4}\,k^3\,H^3\,\left(\frac{H}{m}\right)^{5/2}\left[\left(\frac{m}{H}\right)^{\frac{m^2}{3\,H^2}}-\left(\frac{\Lambda_{IR}}{H}\right)^{\frac{2\,m^2}{3\,H^2}}\right]
\end{align}
where we have assumed $m\ll H$ and used the first of eqs.~(\ref{bogosc}). In eq.~(\ref{pscalmassive}), $\Lambda_{IR}$ corresponds to the scales that left the horizon at the beginning of inflation, so that the total number of efoldings of inflation is given by $N_{\rm Tot}\equiv\log(H/\Lambda_{IR})$. 

Depending on the total duration of inflation, eq.~(\ref{pscalmassive}) simplifies to two different expressions. If $N_{\rm Tot}\gg \frac{3\,H^2}{2\,m^2}$ (the case which includes the limit $\Lambda_{IR}\to 0$), then 
\begin{align}\label{noir}
P_{\cal Q}(k)\simeq -\frac{3\,e^2}{8\,\pi^4}\,k^3\,H^3\,\left(\frac{H}{m}\right)^{5/2}\,.
\end{align}

If, on the contrary, inflation did not last for too long and $N_{\rm Tot}\ll \frac{3\,H^2}{2\,m^2}$ then
\begin{align}\label{yesir}
P_{\cal Q}(k)\simeq -\frac{e^2}{4\,\pi^4}\,k^3\,H^3\,\left(\frac{H}{m}\right)^{1/2}\,\log\left(\frac{m}{\Lambda_{IR}}\right)\,.
\end{align}

Finally, we note that eq.~(\ref{pscalmassive}) was obtained assuming that the dominant contribution to eq.~(\ref{pqscalar}) is given by the regime of integration of lowest $q$, $\Lambda_{IR}\lesssim q\lesssim\sqrt{m\,H}$, i.e., by using the expression for $|\beta_\bq|^2$ given by the first line of eq.~(\ref{bogosc}). However if the scale of interests, characterized by the wave number $k$, are such that $k>\sqrt{m\,H}$, then the scalar field will be effectively massless. The exact Bogolyubov coefficients for a massless scalars read
\begin{align}\label{bogomassless}
\alpha_\bq=\frac{-H^2+2\,i\,q+2\,q^2}{2\,q^2}\,e^{i\,q/H},\quad \beta_\bq=\frac{H^2}{2\,q^2}\,e^{i\,q/H}\,.
\end{align}
Introducing these expressions into eq.~(\ref{pqscalar}) we obtain the simple expression, valid for one massless scalar species
\begin{align}
P_{\cal Q}(k)&=-e^2\,H^4\frac{k^3}{2^5\pi^5}\int \frac{d^3\bq}{q^3\,|\bk+\bq|}\nonumber\\
&\simeq -e^2\,H^4\frac{k^2}{2^3\pi^4}\,\left(N_{\rm Tot}-N_k\right)\,,
\end{align}
where $N_k$ corresponds to the number of efoldings before the end of inflation at which the scale $k$ left the horizon, so that $N_k\simeq 50$.

\section{Effects of the electric field during inflation}%

One might worry that the charge fluctuations generated during inflation produce an electric field which might either oppose further charge separation or annihilate charge via Schwinger pair production. Here we discuss why, in general, this is not the case.

The rate of change of a physical momentum $p$, due to the expansion of the Universe, is given by $H\,p$. For the effect of the electric field to be negligible with respect to that of cosmological expansion we then require $e\, {\cal E}_p\ll H\,p$, where $ {\cal E}_p$ is the typical intensity of the electric field in modes with wavelength larger that $1/p$. In other words, the acceleration due to the electric field should be negligible with respect to the proper deceleration due to the expansion of the Universe. We estimate ${\cal E}_p$ using Gauss's law
\begin{eqnarray}\label{gauss}
{\cal E}_p^2\simeq \langle {\cal E}^2\rangle_p= \int^p\frac{dk}{k^3}\,P_{\cal Q}(k)\,.
\end{eqnarray}
Since sub-horizon charge fluctuations are negligible, we assume $p\lesssim H$, and for fermions we obtain $e\, {\cal E}_p\simeq 3\times 10^{-3}\,e^2\,H^{3/2}\,p^{1/2}$ so that only very low momentum modes with $p\lesssim 10^{-5}\,e^4\,H\simeq 10^{-7}\,H$ are affected by the electric field. Since most of the charge is in modes with $p={\cal O}(m)\gg 10^{-7}\,H$, the effect of the electric field on fermions can be safely neglected.

For scalars things are more complicated. A charged scalar $\phi$ with mass $m\lesssim H$ gets large fluctuations with variance $\langle|\phi|^2\rangle=\frac{3\,H^4}{4\,\pi^2\,m^2}$ and with a correlation length $\sim \int d^3\bk\,k^{-1}\,\left|\phi_k\right|^2$ that is IR-divergent. This implies that $\phi$ acts as a uniform Higgs field, and that the photon gets a mass~\cite{Prokopec:2002jn} $m_\gamma\sim e\,\langle |\phi|^2\rangle^{1/2}\simeq .3\,e\,H^2/m$, which therefore imposes an infrared cutoff in the integral~(\ref{gauss})\footnote{In the case of effectively massless scalars one gets $\langle|\phi|^2\rangle=\frac{H^2}{4\,\pi^2}\,N_{\mathrm Tot}$, so that $m_\gamma\simeq .15\,e\,H\,\sqrt{N_{\rm Tot}}$.}. As a consequence, for $p\lesssim m_\gamma$ the range of integration in eq.~(\ref{gauss}) is vanishing and the electric field is  negligible. On the other hand, the discussion of section~\ref{sub:scalars} above shows that most of the contribution to the electric charge of the Universe comes from the very infrared modes with $p\sim 1/R\ll m_\gamma$. Therefore, the effect of the electric field is negligible.

Another possibility is that the electric field produced by the charge fluctuations ends up annihilating the fluctuations themselves via Schwinger effect. Schwinger pair production is effective if a charged particle $\chi_{\rm Schw}$ with mass $m_{\rm Schw}^2\lesssim e\,\cal E/\pi$ exist, provided the coherence length of the electric field $\lambda$ satisfies $\lambda>2\,\pi\,m_{\rm Schw}/(e\,{\cal E})$~\cite{Dunne:2005sx}. Both conditions give an upper bound on $m_{\rm Schw}$ and must both be satisfied for Schwinger pair production to be effective. 

In the case of fermionic charge, the electric field will have a typical intensity ${\cal E}\sim  3\times 10^{-3}\,e\,H^2$ and its coherence length is approximately $2\,\pi/H$, so that the Schwinger phenomenon is effective if $m_{\rm Schw}\lesssim e\,{\cal E}/H\simeq 3\times 10^{-4}\,H$. 

In the case of charge generated by scalars the coherence length of the electric field is set by the mass of the photon, $\lambda=2\pi/m_\gamma$. As a consequence, if $m_\gamma\gtrsim H$ then the electric field will be negligible, as the infrared cutoff $\sim m_\gamma$ of the electric field is larger than its ultraviolet cutoff $\sim H$ determined by the absence of charge fluctuations at subhorizon scales. The mass of the photon will be larger than $H$ for $m\lesssim .1\,H$. Therefore as long as there is a scalar with mass smaller than $.1\,H$ we should not worry about the Schwinger effect. For scalars with $.1\,H\lesssim m\lesssim H$ we insert eq.~(\ref{pscalmassive}) into eq.~(\ref{gauss}) and take $p\simeq H$ as ultraviolet cutoff. We thus obtain ${\cal E}\simeq .05\,e\,H^2\left(H/m\right)^{5/4}$. By evaluating numerically the condition that $m_{\rm Schw}$ be smaller both than $e\,{\cal E}/m_\gamma$ and than $\sqrt{e\,{\cal E}/\pi}$ we obtain that, for $.1\,H\lesssim m\lesssim H$, the Schwinger effect can be efficient if $m_{\rm Schw}\lesssim .1\,H$. As we stated above, if the field $\chi_{\rm Schw}$ is a scalar, then its large scale fluctuations will contribute to $m_\gamma$ via a Higgs effect, yielding $m_\gamma\gtrsim H$. Therefore, the effect will be important only if $\chi_{\rm Schw}$ is a fermion.

To sum up, Schwinger effect will affect the charge fluctuations only if there exists during inflation a fermion whose mass is smaller than $3\times 10^{-4}\,H$, if the charges originate from the fluctuations of a fermion, or $10^{-1}\,H$, if they originate from a scalar with $.1\,H\lesssim m\lesssim H$. It is worth noting that, since we do not know what is the expectation value of the Higgs field (or of any other scalar field that carries Standard Model charge) during inflation, the mass of the particles of the Standard Model will generally have mass that is different from the one measured today.

\section{Constraints from observations}%

After inflation ends, the charged fermions and scalars considered above will decay into ordinary matter. However, since electric charge is conserved,  the charge density produced during inflation will not be affected. In the post-inflationary Universe, a charge fluctuation is associated to a magnetic field and to vorticity~\cite{Caprini:2003gz} (the electric field being rapidly driven to zero by the large conductivity of the primordial plasma). Observational constraints on vorticity and on the intensity of cosmological magnetic fields impose then an upper bound on the charge density in the Universe. To  see how the values of ${\cal Q}_R$ derived above compare to the constraints of~\cite{Caprini:2003gz}, we define the quantity 
\begin{equation}
y_R=\frac{\sqrt{|\langle {\cal Q}_R^2\rangle|}}{e\,n_B}\,,
\end{equation}
where $n_B$ is the number density of baryons, $n_B\simeq 1.5\times 10^{-10}\, T^3$. The bound~\cite{Caprini:2003gz} depends somehow on the spectral index of the magnetic field, but reads approximately $y_{R=.1\,h^{-1}\,{\mathrm {Mpc}}}\lesssim 10^{-26}$. We will assume, as we did  above, that reheating is instantaneous.

\subsection{Fermions} 

We insert eq.~(\ref{fermion_final}) into eq.~(\ref{qrqk}), we use the fact that ${\cal Q}_R$ scales as the inverse of the volume element, and that $R$ in eq.~(\ref{qrqk}) is a comoving distance. Assuming $g_*\simeq 10^2$ at the time of reheating, setting $T=T_0\simeq 3\times 10^{-4}$~eV, and taking $R\simeq .1\,h^{-1}$~Mpc, we find
\begin{equation}\label{yrfermions}
y_{R=.1\,h^{-1}\,{\mathrm {Mpc}}}\simeq 3\times 10^{-33}\,\sqrt{\frac{f^f(m/H)}{10^{-5}}}\,\left(\frac{H}{9\times 10^{13}\,{\mathrm {GeV}}}\right)^{3/4}\,,
\end{equation}
where we have normalized $H$ to its maximum possible value, that is determined by the non-observation of tensor modes in the CMB. Fermions fall short of the constraint by at least 8 orders of magnitude.

\subsection{Scalars}

For massive scalars with a ``long'' inflation ($N_{\rm Tot}\gg H^2/m^2$), an analogous computation yields
\begin{align}\label{yrscalars}
y_{R=.1\,h^{-1}\,{\mathrm {Mpc}}}\simeq 4\times 10^{-32}\,\left(\frac{H}{9\times 10^{13}\,{\mathrm {GeV}}}\right)^{3/4}\left(\frac{H}{m}\right)^{5/4}\,,
\end{align}
so that the constraint $y_{R=.1\,h^{-1}\,{\mathrm {Mpc}}}\lesssim 10^{-26}$ is satisfied unless the scalar is very light $m\lesssim 5\times 10^{-5}\,H$. For these small values of $m$, however, the condition $N_{\rm Tot}\gg H^2/m^2$ is easily violated, and we should rather use eq.~(\ref{yesir}) to compute $y_R$, yielding the bound $m\gtrsim 10^{-23}\,\log^2(m/\Lambda_{IR})\,H$. If this bound is violated, however, $m$ will be so small that it is more natural to consider the exactly massless case, that gives
\begin{align}\label{yrmasslessscal}
y_{R=.1\,h^{-1}\,{\mathrm {Mpc}}}\simeq 10^{-20}\,\left(\frac{H}{9\times 10^{13}\,{\mathrm {GeV}}}\right)\sqrt{N_{\rm Tot}}\,,
\end{align}
that, for the maximal allowed value of $H$, exceeds the observational limit by at least $7$ orders of magnitude even in the case of short inflation, $N_{\rm Tot}={\cal O}(10^2)$.

\section{Conclusions}%

If one or more electrically charged species have a mass smaller than the Hubble parameter during inflation, then our Universe will typically carry a net electric charge. We have found that each species whose mass is of the order of the Hubble parameter contributes a charge density that is $5\div 7$ order of magnitude below the observational limits. Very light scalars, however, can contribute much more charge density, and in the limit of massless scalars the resulting charge density can exceed by seven (or more, depending on the duration of inflation) orders of magnitude the constraints of~\cite{Caprini:2003gz}, unless the Hubble parameter during inflation is well below the ``high scale inflation'' regime $H\simeq 10^{13}$~GeV.

We should point out that our analysis concerns the simpler regime of constant mass particles during inflation with constant Hubble parameter. While it is straightforward to extend our conclusions to the case of adiabatically evolving $m$ or $H$, it would be especially interesting consider the case where the parameters in the theory are rapidly evolving, for instance as a consequence of a phase transition. Finally, it would be interesting to study whether charged scalars that are experiencing a period of tachyonic evolution can generate large charge fluctuations. 

While in this paper we have focused on electric charge,  our arguments are valid for any conserved quantum number. In particular, they will be unchanged if we replace electric charge by baryon (this was the case examined in~\cite{Kobakhidze:2015zka}) or lepton number, with the simplification that such charges are not associated to any long range forces. We hope to return to this subject in a future work.

{\em Acknowledgments.---}
We thank Chiara Caprini, John Donoghue, Nemanja Kaloper, Marco Peloso, Arttu Rajantie and Jennie Traschen for useful discussion. This work has been partially supported by the NSF grants PHY-1205986 and PHY-1520292.

\begin{appendix}

\section{Bogolyubov coefficients}\label{app:bogos}

To properly motivate our procedure of normal ordering/renormalization, let us examine the structure of the Hamiltonian for a complex scalar field. We start by writing the action of the scalar in terms of the comoving field as
\begin{equation}
S = \int d^4 \textbf{x} \left\{ \varphi' {\varphi^*}' + \varphi^* \left[ \Delta + \frac{a''}{a} - a^2 m^2 \right]\varphi \right\},
\end{equation}
where the conjugate momenta of $\varphi$ and $\varphi^*$ are $\Pi_\varphi = {\varphi^*}'$ and $\Pi_{\varphi^*} = {\varphi}'$.

Using eq.~(\ref{scalardeco}), the Hamiltonian can then be written as
\begin{align} \label{nonadham}
H = \int d^3 k &\left[ \left( a_\textbf{k} a^\dagger_\textbf{k} + b^\dagger_{-\textbf{k}} b_{-\textbf{k}} \right) g(\bk,\,\tau)\right. \nonumber\\
&+\left.a_\textbf{k} b_{-\textbf{k}} f(\bk,\,\tau) + b^\dagger_{-\textbf{k}} a^\dagger_{\textbf{k}} f^*(\bk,\,\tau) \right]\,,
\end{align}
where
\begin{align}
&f(\bk,\tau) = {\phi'(\bk)}^2 + \omega_k^2 \, \phi(\bk)^2\,,\nonumber\\
&g(\bk,\tau) = |\phi'(\bk)|^2\, +\omega_k^2\, |\phi(\bk)|^2\,.
\end{align}

The Hamiltonian in the above form is not diagonal and the definition of the number operator is unclear since the $a$ and $b$ operators do not annihilate energy eigenstates.  We can however diagonalize the above Hamiltonian by performing a Bogolyubov transformation on the operators,
\begin{equation} \label{bogo}
\left( \begin{matrix} \tilde a_{\textbf{k}}(\tau) \\ \tilde b^\dagger_{-\textbf{k}}(\tau) \end{matrix} \right) = \left( \begin{matrix} \alpha_\bk(\tau) & \beta^*_\bk(\tau) \\ \beta_\bk(\tau)&\alpha^*_\bk(\tau) \end{matrix} \right) \left( \begin{matrix} a_{\textbf{k}} \\ b^\dagger_{-\textbf{k}} \end{matrix} \right),
\end{equation}
where $\alpha_\bk$ and $\beta_\bk$ are the Bogolyubov coefficients, and $\tilde{a}_\textbf{k}$ and $\tilde{b}_\textbf{k}$ are annihilation operators with an associated vacuum, $\ket{\tilde{0}}$. By imposing that both the $a_\bk$, $b_{-\bk}$ and the $\tilde{a}_\bk$, $\tilde{b}_{-\bk}$ operators satisfy canonical commutation relations, we find the constraint $|\alpha_\bk|^2-|\beta_\bk|^2=1$. We can invert the transformation~(\ref{bogo}) in terms of the original operators as
\begin{equation}
\left( \begin{matrix} a_{\textbf{k}}(\tau) \\ b^\dagger_{-\textbf{k}}(\tau) \end{matrix} \right) = \left( \begin{matrix} \alpha^*_\bk(\tau) & - \beta^*_\bk(\tau) \\ - \beta_\bk(\tau)&\alpha_\bk(\tau)\end{matrix} \right) \left( \begin{matrix} \tilde{a}_{\textbf{k}} \\ \tilde{b}^\dagger_{-\textbf{k}} \end{matrix} \right),
\end{equation}
which allows us to write the Hamiltonian compactly as
\begin{equation} \label{hamdia}
H = \int d^3 \textbf{k} \; \omega_\textbf{k} \left[ \tilde{a}_\textbf{k} \,\tilde{a}^\dagger_\textbf{k} + \tilde{b}^\dagger_\textbf{k} \,\tilde{b}_\textbf{k}  \right],
\end{equation}
where $f(\bk,\,\tau)$ and $g(\bk,\,\tau)$ must satisfy
\begin{equation} \label{fandg}
f(\bk,\,\tau) = 2\,  \omega_k\,\alpha_\bk\, \beta_\bk \,,\quad g(\bk,\,\tau) = \omega_k \,\left( |\alpha_\bk|^2 + |\beta_\bk|^2 \right),
\end{equation}
in order for eq.~(\ref{hamdia}) to hold. In contrast to the Hamiltonian in terms of the original operators $a_\textbf{k}$ and $b_\textbf{k}$, for the Hamiltonian in eq.~(\ref{hamdia}) we can unambiguously define the number operators, $\tilde{N}^a_\textbf{k} = \tilde{a}^\dagger_\textbf{k}\, \tilde{a}_\textbf{k}$ and $\tilde{N}^b_\textbf{k} = \tilde{b}^\dagger_\textbf{k}\, \tilde{b}_\textbf{k}$ which will tell us the number of quanta of the energy eigenstates of $\varphi$ in the states whose vacuum is $\ket{\tilde{0}}$.  We must also normal order (with respect to the tilded  operators) so that the energy associated with the vacuum $|\tilde{0}\rangle$ vanishes, 
\begin{equation}
\Braket{\tilde{0} | : H : | \tilde{0}} = \int d^3 \textbf{k} \; \omega_\textbf{k} \; \Braket{\tilde{0} | \tilde{a}^\dagger_\textbf{k} \tilde{a}_\textbf{k} + \tilde{b}^\dagger_\textbf{k} \tilde{b}_\textbf{k} | \tilde{0}} = 0.
\end{equation}

However, if we act on the original vacuum $\ket{0}$, with the adiabatic number operators we find
\begin{equation} \label{ham}
\Braket{0 | \tilde{N}^a_\textbf{k}  | 0} = \Braket{0 | \tilde{N}^b_\textbf{k}  | 0} = |\beta_\bk|^2 \neq 0,
\end{equation}
thus if $|\beta_k|^2 \neq 0 $ this is interpreted as particle creation.  The Hamiltonian remains diagonalized when eqs.~(\ref{fandg}) are satisfied and only in this case can we define the number of quanta of $\varphi$ in a meaningful and unambiguous manner.  We can ensure the relations in eq.~(\ref{fandg}) are obeyed if the mode functions satisfy a WKB-type solution,
\begin{align} \label{adi}
&\varphi(\bk,\tau) = \alpha_\bk\, \tilde{\varphi} (\bk,\tau) + \beta_\bk\, \tilde{\varphi}^* (\bk,\tau) \,,\nonumber\\
&\tilde{\varphi}(\bk,\tau) \equiv \frac{1}{\sqrt{2\omega_k}} e^{-i \int \omega_k d \tau},
\end{align}
with the condition for adiabaticity $\frac{\omega'}{\omega^2} \ll 1$.

To summarize, an adiabatic vacuum exists (along with its associated adiabatic operators) for early times during which our Hamiltonian is diagonalized allowing us to clearly define the number of particles\footnote{We assume that the initial charge density vanishes, but even if there is an initial density it will quickly be diluted by the de Sitter expansion}.  Subsequently, a period of nonadiabacity occurs and our Hamiltonian has the nontrivial form of eq.~(\ref{nonadham}).  After this period of nonadiabacity our system evolves adiabatically again and we can once again unambiguously define the number of particles.  The Bogolyubov transform in eq.~(\ref{bogo}) allows us to relate these three regions, and the Bogolyubov coefficients enables us to calculate the number of particles created during the period of nonadiabaticity.  Finally, we must normal order the tilded operators (which are the operators that allow to count our particles), but calculate expectation values using $\ket{0}$ since this is the state in which particle creation takes place.

\section{Calculation of Bogolyubov Coefficients}\label{app:bogocalc}

The exact solution for the mode functions of a scalar field of mass $m$  during inflation with Hubble parameter $H$ is
\begin{equation}
\varphi_I = \sqrt{-k \tau} \left[ A_k \mathrm{H}_\nu^{(1)}( - k \tau) + B_k \mathrm{H}_\nu^{(2)}( - k \tau) \right], 
\end{equation}
where $A_k$ and $B_k$ are arbitrary constants.  The adiabatic solution to the mode functions is
\begin{align} \label{WKB} 
&\varphi^{\mathrm{WKB}}(\bk,\tau) = \alpha_\bk\, \tilde{\varphi} (\bk,\tau) + \beta_\bk\, \tilde{\varphi}^* (\bk,\tau)\,,\nonumber\\
&\tilde{\varphi}(\bk,\tau) \equiv \frac{1}{\sqrt{2\omega_k}} e^{-i \int \omega_k d \tau}\,. 
\end{align}

If our initial state contains no particles, then $\alpha_\bk^{IN} = 1$ and $\beta_\bk^{IN} = 0$, which leads to
\begin{equation}
A_k = \sqrt{ \frac{\pi}{4k} }\,,\quad B_k = 0.
\end{equation}

The adiabatic condition ($\frac{\omega'}{\omega^2} \ll 1$) at the end of inflation reads $m\gg a'/a^2$, where we have assumed that we are looking at long wavelength modes for which $\omega\simeq m\,a$. For masses on the order of $H$ or less, the modes are not evolving adiabatically at the end of inflation, and the number of particles is therefore not a well-defined quantity.  We can however join the end of inflation to a radiation epoch with $a=H\,\tau+2$, where the adiabatic condition reads
\begin{equation}
\frac{\omega'}{\omega^2} = \frac{H m^2 (H \tau +2)}{\left(m^2 (H \tau +2)^2+k^2\right)^{3/2}} \ll 1,
\end{equation}
showing that a well defined concept of particle will exist assuming we wait long enough, $ \tau \gg \frac{1}{\sqrt{m H}}$. 

The equation of motion of a massive scalar during the radiation epoch is
\begin{equation}
\varphi_R'' + \left( k^2 + m^2 ( H\tau + 2)^2 \right) \varphi_R = 0,
\end{equation}
whose solution can be written in terms of  parabolic cylinder functions as
%
\begin{widetext}
\begin{equation}
\varphi_R(\tau) = a_k \,D_{ -\frac{1}{2}- i \frac{k^2}{2\,H\,m} } \left( e^{i\frac{\pi}{4}} \sqrt{\frac{2m}{H}}\left( H\, \tau + 2 \right) \right) + b_k\, D_{ -\frac{1}{2}+ i \frac{k^2}{2\,H\,m} }\left( e^{i\frac{3\,\pi}{4}} \sqrt{\frac{2m}{H}}\left( H\, \tau + 2 \right) \right)\,.
\end{equation}
The constants $a_k$ and $b_k$ are determined by joining the exact solutions during inflation to those during the radiation dominated era, that is, by imposing $\varphi_I(\tau_R) = \varphi_R(\tau_R)$ and  $ \varphi'_I(\tau_R) = \varphi'_R(\tau_R)$, where $\tau_R=-1/H$ denotes the time of the end of inflation.  The adiabatic solution for the mode functions after inflation will have the form of eq.~(\ref{WKB}) with $\omega_k^2 = k^2 +m^2(H\tau + 2)^2$.  We can solve for the Bogolyubov coefficients by matching the exact solution to the adiabatic solution for late times ($\tau \rightarrow + \infty$)
\begin{align}
&\varphi_R(\tau \rightarrow +\infty) \approx \frac{ e^{-\frac{\pi  \tilde{k}^2}{8\tilde{m} }}}{\sqrt{\tilde{\tau} \sqrt{2 \tilde{m} }}} \left[a_k e^{\frac{\pi \tilde{k}^2}{4 \tilde{m} }} + b_k \frac{\sqrt{2\,\pi }\,e^{i\pi/4} }{\Gamma \left(\frac{1}{2} -i\frac{\tilde{k}^2}{2 \tilde{m} }\right)}\right]  e^{-\frac{i}{2} \tilde{m}  \tilde{\tau}^2} \tau^{-i\frac{\tilde{k}^2}{2 \tilde{m} }} + b_k \frac{ e^{-\frac{3 \pi  \tilde{k}^2}{8 \tilde{m} }}}{\sqrt{\tilde{\tau} \sqrt{2 \tilde{m}}} }  e^{\frac{i}{2} \tilde{m} \tilde{\tau}^2} \tilde{\tau}^{i\frac{\tilde{k}^2}{2 \tilde{m} }}\,,\nonumber\\
&\varphi^{\mathrm{WKB}}(\tau \rightarrow \infty) \approx \frac{\alpha_k/\sqrt{H}}{\sqrt{2 \tilde{m} \tilde{\tau}}} e^{-\frac{i}{2}\tilde{m} \tilde{\tau}^2} \tau^{-i\frac{\tilde{k}^2}{2 \tilde{m} }} + \frac{\beta_k/\sqrt{H}}{\sqrt{2 \tilde{m} \tilde{\tau}}} e^{\frac{i}{2}\tilde{m} \tilde{\tau}^2} \tau^{i\frac{\tilde{k}^2}{2 \tilde{m} }}\,,
\end{align}
\end{widetext}
%
obtaining
\begin{align}
&\alpha_\bk = \sqrt{H}\left(2\,\tilde{m}\right)^{1/4} e^{\frac{\pi \tilde{k}^2}{8\tilde{m}}} \left(a_k+ b_ke^{- \frac{\pi \tilde{k}^2}{8\tilde{m}}}  \frac{\sqrt{2\,\pi}\,e^{i\pi/4} }{\Gamma \left(\frac{1}{2} -i\frac{ \tilde{k}^2}{2\tilde{m}}\right)}\right) \,,\nonumber\\
&\beta_\bk =\sqrt{H}\left(2\,\tilde{m}\right)^{1/4} e^{-\frac{3\pi \tilde{k}^2}{8\tilde{m}}}\, b_k\,,
\end{align}
where for notational simplicity $\tilde{k} = k/H$, $ \tilde{m} = m/H$ and $\tilde{\tau}=H\,\tau$.  We will be interested in nonrelativistic ($k \ll m a$), superhorizon ($-k \tau_R \ll 1$) modes, for which we find
\begin{equation}
\beta_\bk \approx -e^{i\frac{\pi}{8}}\frac{\Gamma(1/4)}{2\sqrt{2\pi}}\left(\frac{H}{m}\right)^{1/4} \left( \frac{k}{H} \right)^{-\sqrt{9/4-m^2/H^2}}\,,
\end{equation}
where $\Gamma(1/4)/(2\sqrt{2\pi})\simeq .72$. Since $\beta_k$ is non-zero we interpret this as the de Sitter expansion causing quanta of $\varphi$ to be created.  

An analogous study can be performed in the case of massless scalars, and the exact Bogolyubov coefficients take a much simpler form
\begin{align} \label{bogomassless}
&\alpha_\bk = e^{\frac{i k}{H}} \left( 1 + i\,\frac{H}{k}- \frac{H^2}{2k^2} \right)\,,\nonumber\\
&\beta_\bk = e^{\frac{i k}{H}} \frac{H^2}{2 k^2}\,.
\end{align}

\end{appendix}


\begin{thebibliography}{99}

\bibitem{pdg} K. Nakamura et al. (Particle Data Group), JPG 37, 075021 (2010)

\bibitem{Caprini:2003gz} 
  C.~Caprini and P.~G.~Ferreira,
  JCAP {\bf 0502}, 006 (2005)
  [hep-ph/0310066].

\bibitem{Orito:1985cf} 
  S.~Orito and M.~Yoshimura,
  Phys.\ Rev.\ Lett.\  {\bf 54}, 2457 (1985).

\bibitem{Masso:2002vh} 
  E.~Masso and F.~Rota,
  Phys.\ Lett.\ B {\bf 545}, 221 (2002)
  [astro-ph/0201248].
  
\bibitem{Lyttleton:1959zz} 
  R.~A.~Lyttleton and H.~Bondi,
  Proc.\ Roy.\ Soc.\ Lond.\ A {\bf 252}, 313 (1959).
   
\bibitem{alfven}
Alfv\'en, H. and Klein, O. (1962) Arkiv for Fysik, 23, 187-194.

\bibitem{Kaloper:2009nc} 
  N.~Kaloper and A.~Padilla,
  JCAP {\bf 0910}, 023 (2009)
  [arXiv:0904.2394 [astro-ph.CO]].

\bibitem{Ignatiev:1978xj} 
  A.~Y.~Ignatiev, V.~A.~Kuzmin and M.~E.~Shaposhnikov,
  Phys.\ Lett.\ B {\bf 84}, 315 (1979).
  
\bibitem{Langacker:1980kd} 
  P.~Langacker and S.~Y.~Pi,
  Phys.\ Rev.\ Lett.\  {\bf 45}, 1 (1980).
  
\bibitem{Enqvist:1988br} 
  K.~Enqvist, K.~W.~Ng and K.~A.~Olive,
  UMN-TH-656/88.
  
\bibitem{Dolgov:1990sg} 
  A.~D.~Dolgov,
  Phys.\ Lett.\ B {\bf 276}, 347 (1992).
  
\bibitem{Dolgov:1993mu} 
  A.~Dolgov and J.~Silk,
  Phys.\ Rev.\ D {\bf 47}, 3144 (1993).
  
\bibitem{Dolgov:2006bc} 
  A.~Dolgov and D.~N.~Pelliccia,
  Phys.\ Lett.\ B {\bf 650}, 97 (2007)
  [hep-ph/0610421].
  
\bibitem{Dubovsky:2000av} 
  S.~L.~Dubovsky, V.~A.~Rubakov and P.~G.~Tinyakov,
  JHEP {\bf 0008}, 041 (2000)
  [hep-ph/0007179].

\bibitem{Calzetta:1997ku} 
  E.~A.~Calzetta, A.~Kandus and F.~D.~Mazzitelli,
  Phys.\ Rev.\ D {\bf 57}, 7139 (1998)
  [astro-ph/9707220].
  
\bibitem{Giovannini:2000dj} 
  M.~Giovannini and M.~E.~Shaposhnikov,
  Phys.\ Rev.\ D {\bf 62}, 103512 (2000)
  doi:10.1103/PhysRevD.62.103512
  [hep-ph/0004269].

\bibitem{D'Onofrio:2012qy} 
  M.~D'Onofrio, R.~N.~Lerner and A.~Rajantie,
  JCAP {\bf 1210}, 004 (2012)
  [arXiv:1207.1063 [astro-ph.CO]].

\bibitem{Kobakhidze:2015zka} 
  A.~Kobakhidze and A.~Manning,
  Phys.\ Rev.\ D {\bf 91}, no. 12, 123529 (2015)
  [arXiv:1506.01511 [hep-ph]].
  
\bibitem{Nilles:2001fg} 
  H.~P.~Nilles, M.~Peloso and L.~Sorbo,
  JHEP {\bf 0104}, 004 (2001)
  [hep-th/0103202].
  
\bibitem{Prokopec:2002jn} 
  T.~Prokopec, O.~Tornkvist and R.~P.~Woodard,
  Phys.\ Rev.\ Lett.\  {\bf 89}, 101301 (2002)
  [astro-ph/0205331].
  
\bibitem{Dunne:2005sx} 
  G.~V.~Dunne and C.~Schubert,
  Phys.\ Rev.\ D {\bf 72}, 105004 (2005)
  [hep-th/0507174].

\end{thebibliography}
\end{document}